\documentclass[aps,prc,superscriptaddress,nofootinbib,floatfix,twocolumn]{revtex4-1}
\usepackage{lineno}
\usepackage[colorlinks,linkcolor=blue,citecolor=blue,filecolor=black,urlcolor=blue]{hyperref}
\usepackage{epsfig,graphics}
\usepackage{graphicx}
\usepackage{bm}
\usepackage{amssymb}
\usepackage{amsmath}
\usepackage{multirow}
\usepackage{MnSymbol}
\usepackage{ragged2e}
\usepackage{moresize}
\usepackage{enumitem}
\makeatletter 
\@addtoreset{equation}{section}
\makeatother  

\renewcommand{\figurename}{\textbf{Fig.}}

\usepackage{caption}
\DeclareCaptionLabelFormat{nature}{\textbf{#1\ #2}\ $\vert$\ }
\newcommand{\lr}[1]{\left\langle #1\right\rangle}
\newcommand{\pT} {\ensuremath{p_{\mathrm{T}}}}
\newcommand{\snn}{\mbox{$\sqrt{s_{\mathrm{NN}}}$}}
\newcommand{\nch} {N_{\mathrm{ch}}}
\newcommand{\Deta}{\Delta \eta}
\newcommand{\etaref}{\eta_{\mathrm{ref}}}
\newcommand{\auau}{$^{197}$Au+$^{197}$Au}
\newcommand{\uuuu}{$^{238}$U+$^{238}$U}
\begin{document}

\title{Longitudinal structure of the quark--gluon plasma from differently shaped nuclei}
\author{Jiangyong Jia}\email{jiangyong.jia@stonybrook.edu}\affiliation{\sbu}
\author{Chunjian Zhang}\email{chunjianzhang@fudan.edu.cn}\affiliation{\fdu}\affiliation{\moe}
\author{Shengli Huang}\email{shengli.huang@stonybrook.edu}\affiliation{\sbu}
\newcommand{\sbu}{Department of Chemistry, Stony Brook University, Stony Brook, NY 11794, USA}
\newcommand{\bnl}{Physics Department, Brookhaven National Laboratory, Upton, NY 11976, USA}
\newcommand{\moe}{Key Laboratory of Nuclear Physics and Ion-beam Application (MOE), and Institute of Modern Physics, Fudan University, Shanghai 200433, China}
\newcommand{\fdu}{Shanghai Research Center for Theoretical Nuclear Physics, NSFC and Fudan University, Shanghai 200438, China}
\begin{abstract}
Ultrarelativistic collisions of atomic nuclei produce the quark--gluon plasma (QGP), an extremely hot, dense state of matter. The QGP behaves like a nearly perfect fluid, so its final-state momentum distributions can be inverted to reveal its initial-state geometry. This programme has succeeded in the transverse plane but made less progress along the beam, where short-range nonflow correlations mask the longitudinal signal. Anisotropic flow has been successfully used to image the shapes of the colliding nuclei; here we use nuclear shape to image the QGP's 3D geometry---the same idea in reverse. We show in simulations that the nonflow can be removed by comparing collisions of nuclei with similar masses but different shapes. We find that nuclear deformation changes the magnitude of the elliptic flow but not its longitudinal profile, so the shape difference recovers the full longitudinal structure. The predicted two-particle decorrelation map reveals both a highly non-linear rapidity dependence inaccessible to conventional observables, and an unquantified bias in standard flow measurements themselves. Our strategy extends to the longitudinal dependence of the QGP's triangularity and size. Nuclear shape thus becomes a tool for 3D imaging of the QGP and the quantum fluctuations of the nuclear wavefunctions that seed it.
\end{abstract}
\maketitle
Atomic nuclei are many-body quantum systems of protons and neutrons, themselves built from quarks and gluons bound by the strong force of quantum chromodynamics (QCD). Ultrarelativistic collisions liberate these constituents into the QGP, the state of matter that filled the Universe within the first few microseconds after the Big Bang. Despite its transience (lifetime $\sim 10^{-23}$~s) and microscopic extent ($\sim 10^{-14}$~m), the QGP behaves like a near-perfect fluid, with the smallest ratio of shear viscosity to entropy density of any known substance~\cite{Heinz:2013wva,Romatschke:2017ejr,Busza:2018rrf,Everett:2020xug}. This fluidity translates the initial spatial distribution of quarks and gluons into correlated momentum anisotropies among the thousands of emitted particles, generating a collective velocity field known as anisotropic flow~\cite{Ollitrault:2023wjk}.

\begin{figure*}[htbp]
\centering
\includegraphics[width=0.8\linewidth]{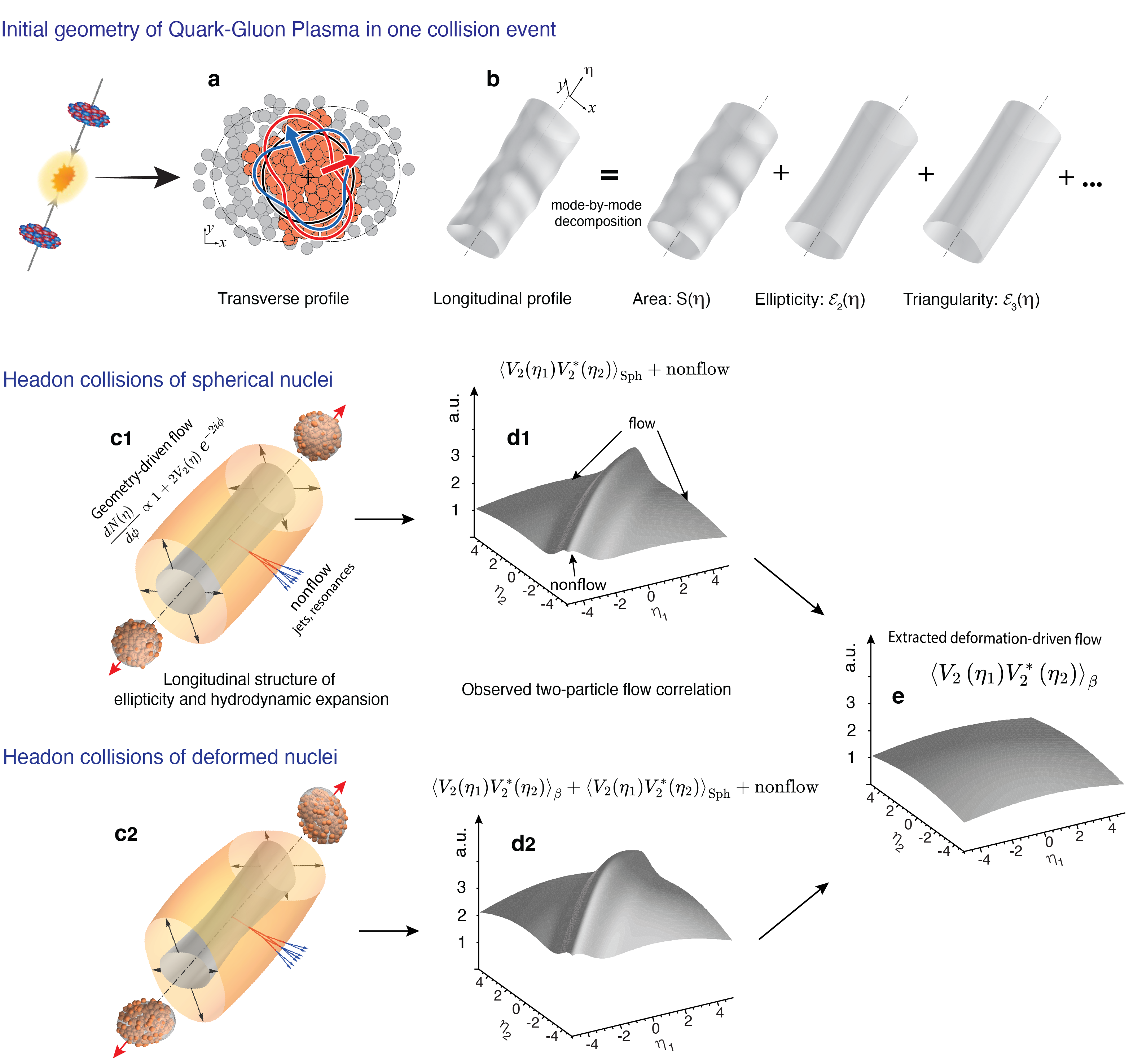}
\caption{\label{fig:cartoon}\textbf{From nuclear shape to longitudinal flow correlations.} (\textbf{a},\textbf{b}) Initial collision geometry in a single event: the transverse ($xy$) distribution of participating (orange) and spectator (grey) nucleons, with the overlap region decomposed into a mean size and shape components, the arrows giving their orientations (\textbf{a}); and the longitudinal profile of that region (\textbf{b}), decomposed along pseudorapidity $\eta=-\ln\tan(\theta/2)$ ($\theta$ is the polar angle to the beam) into the overlap area $S(\eta)$, ellipticity $\mathcal{E}_2(\eta)$, triangularity $\mathcal{E}_3(\eta)$ and higher-order components. (\textbf{c}) Head-on collisions of spherical (\textbf{c1}) and prolate-deformed (\textbf{c2}) nuclei, elliptic component only. The inner grey surface is the initial ellipticity $\mathcal{E}_2(\eta)$, interpolating between the projectile and target shapes. The outer orange surface represents the elliptic flow $V_2(\eta)$ modulation generated by hydrodynamic pressure gradients. (\textbf{d}) The resulting event-averaged two-particle correlations $\lr{V_2(\eta_1)V_2^{*}(\eta_2)}$: a fluctuation-driven baseline plus a short-range nonflow ridge from jets and decays near $\eta_1\approx\eta_2$ for spherical nuclei (\textbf{d1}), with an additional deformation-driven component for deformed nuclei (\textbf{d2}). (\textbf{e}) The difference \textbf{d2}$-$\textbf{d1}, which isolates the deformation-driven correlation over the full rapidity range. Panels (\textbf{b})--(\textbf{d}) run from the geometry to the correlation. The measurement runs the other way, and panel (\textbf{e}) makes that inversion possible by removing the nonflow.}
\vspace*{-0.5cm}
\end{figure*}

The collective expansion of the QGP is well described by relativistic viscous hydrodynamics. In each collision event, quantum fluctuations in the nuclear wavefunctions and the stochastic nature of energy deposition produce a QGP fireball with a unique size and a unique set of elliptic, triangular, and higher-order shape components (Fig.~\ref{fig:cartoon}a). Hydrodynamic evolution faithfully converts these spatial anisotropies into momentum anisotropies of the emitted particles. This correspondence can be inverted: the measured momentum distributions reveal how quarks and gluons were distributed at the moment of impact, and hence the wavefunctions of the colliding nuclei~\cite{Niemi:2012aj,Jia:2022ozr}. In recent years, this programme has successfully constrained the transverse geometry of the QGP and, by extension, the structure of atomic nuclei---their radii, skin thickness, and intrinsic deformations~\cite{Jia:2021oyt,Xu:2021vpn,Giacalone:2023hwk,STAR:2024eky}.

In contrast, the structure of the QGP along the beam direction, an equally fundamental aspect of the collision, remains much less explored. Energy from individual nucleon--nucleon interactions is deposited into QCD strings whose starting positions and lengths fluctuate strongly along pseudorapidity $\eta$~\cite{Shen:2017bsr}. As a result, the size and shape of the fireball become fluctuating functions of rapidity (Fig.~\ref{fig:cartoon}b). They interpolate between the shape of the projectile nucleus at far-forward rapidity and that of the target nucleus at far-backward rapidity~\cite{Bozek:2015bna} (Fig.~\ref{fig:cartoon}c).  The same hydrodynamic response acts at every rapidity, so correlations between particles emitted at different rapidities encode the corresponding correlations of the initial geometry~\cite{Bozek:2010vz,Jia:2014ysa,Pang:2015zrq,Wu:2018cpc,Shen:2017bsr}. In particular, flow vectors at well-separated rapidities gradually lose their mutual alignment in both magnitude and orientation, a phenomenon known as flow decorrelation. Mapping this longitudinal structure would complete the three-dimensional imaging of the QGP, constraining the rapidity profile of energy deposition, evolution of quark and gluon distributions along rapidity, and quantum fluctuations of the nuclear wavefunctions.

In practice, however, this mapping is hindered by short-range nonflow correlations, created at late times by jet fragmentation and resonance decays~\cite{Feng:2024eos,Wang:2026hak}. They appear as a pronounced ridge on top of the genuine long-range correlations (Fig.~\ref{fig:cartoon}d) and overwhelm the decorrelation signal~\cite{Voloshin:2008dg,Xu:2020koy}. Existing approaches mitigate nonflow with ratio observables $r_n(\eta)$, constructed using reference particles at large rapidity~\cite{CMS:2015xmx,ATLAS:2017rij,ATLAS:2020sgl,ATLAS:2023rbh,ALICE:2023tvh}. As we show below, these provide only indirect and biased information on the genuine longitudinal correlations. To date, no established experimental method exists to recover the full longitudinal structure of the QGP.

{\bf Nuclear shape as an instrument.} We overcome the nonflow limitation by comparing nuclei of similar mass but different shape. Previous studies have used anisotropic flow to image the intrinsic deformations of nuclei such as $^{238}$U, $^{129}$Xe and $^{20}$Ne~\cite{STAR:2024eky,STAR:2025elk,ATLAS:2022dov,Bally:2021qys,Zhang:2021kxj,ATLAS:2025nnt,ALICE:2025luc,CMS:2025tga}. This paper carries out the same idea in reverse: instead of using the plasma to measure nuclear shape, we use nuclear shape difference to remove the nonflow and thereby recover the full longitudinal structure of the plasma.

This strategy is illustrated in Fig.~\ref{fig:cartoon}. Most atomic nuclei are not spherical but carry an intrinsic quadrupole deformation, quantified by the parameter $\beta$. In head-on collisions of near-spherical $^{197}$Au (Fig.~\ref{fig:cartoon}c1), the fireball ellipticity arises almost entirely from random fluctuations of nucleon positions, whereas for prolate $^{238}$U (Fig.~\ref{fig:cartoon}c2) it receives additional contributions from the quadrupole deformation~\cite{Filip:2009zz,Jia:2021qyu}. The short-range nonflow correlations, by contrast, are nearly identical in the two systems. Subtracting the measured correlators (Fig.~\ref{fig:cartoon}d2 minus Fig.~\ref{fig:cartoon}d1) cancels both the nonflow and the fluctuation-driven baseline. This subtraction isolates the deformation-driven correlation over the full rapidity range (Fig.~\ref{fig:cartoon}e), and hence will enable measurements of flow decorrelations free of nonflow. 

Crucially, the fluctuation-driven and deformation-induced components of the geometry originate from the same participating nucleons, and we find that they decorrelate identically along rapidity. This universality is what allows the deformation-driven correlation to image the full flow decorrelation.

{\bf Removing the nonflow.} We now formalize the strategy outlined in Fig.~\ref{fig:cartoon}, using elliptic flow as illustration. The transverse shape of the fireball at each rapidity is quantified by its overlap area $S(\eta)$ and by eccentricity vectors $\mathcal{E}_n(\eta)$ (Fig.~\ref{fig:cartoon}b). The leading ones are the ellipticity $\mathcal{E}_2(\eta)= \varepsilon_2(\eta) e^{2i\Phi_2(\eta)}$ and the triangularity $\mathcal{E}_3(\eta)$. Hydrodynamic evolution maps each mode onto a corresponding flow vector via a linear response, $V_n(\eta) \propto \mathcal{E}_n(\eta)$, which relates both their amplitudes and their phases~\cite{Teaney:2010vd,Gardim:2011xv,Niemi:2012aj}. We focus on the elliptic flow $V_2(\eta)=v_2(\eta)e^{2i\Psi_2(\eta)}$, the response to the elliptically-shaped overlap region~\cite{Busza:2018rrf}, which fluctuates along $\eta$ within a single collision event~\cite{Jia:2014ysa,Pang:2015zrq} (Fig.~\ref{fig:cartoon}c).

Experimentally, this longitudinal structure is accessed through the two-particle correlation (2PC) $V_{2\Delta}(\eta_1,\eta_2)=\lr{V_2(\eta_1)V_2^*(\eta_2)}$. The field $V_2(\eta)$ varies from one event to the next, so $V_{2\Delta}$ is its covariance along rapidity. Measuring the full map therefore fixes the entire second-order structure of the event-by-event distribution of $V_2(\eta)$. This is the sense in which the map images the longitudinal geometry (Extended Data~\ref{app:J}). Decorrelations appear as deviations of the factorization ratio from unity:
\vspace*{-0.2cm}
\begin{align}\label{eq:1}
R_2(\eta_1,\eta_2) = \frac{V_{2\Delta}(\eta_1,\eta_2)}{\sqrt{V_{2\Delta}(\eta_1,\eta_1)V_{2\Delta}(\eta_2,\eta_2)}}\leq 1
\end{align}

Although a simplified version of $R_2(\eta_1,\eta_2)$ was proposed some time ago~\cite{Bozek:2010vz}, it has never been directly measured due to large nonflow correlations at small relative rapidity $|\Deta|\lesssim 1$--2 ($\Deta= \eta_1-\eta_2$) in Fig.~\ref{fig:cartoon}d. Instead, flow decorrelations have been accessed {\it indirectly} in limited rapidity ranges, often by correlating reference particles at forward or backward rapidity (e.g., $\etaref\lesssim-4$) with particles of interest at mid-rapidity (e.g., $0<\eta<2$), using~\cite{CMS:2015xmx}:
\vspace*{-0.2cm}
\begin{align}\label{eq:2}
r_2(\eta) = \frac{V_{2\Delta}(\eta,\etaref)}{V_{2\Delta}(-\eta,\etaref)} \equiv \frac{R_2(\eta,\etaref)}{R_2(-\eta,\etaref)}\;,
\end{align}
where the equivalence holds for symmetric collision systems. However, $r_2$ is a {\it ratio of decorrelations} rather than the decorrelation itself, and hence carries its own biases. It reduces to the true decorrelation only when $\etaref=-\eta$, i.e., $r_2(\etaref) = R_2(\etaref,-\etaref)$.

Although $r_2$ is known to be sensitive to nuclear shape~\cite{Nie:2022gbg}, that sensitivity has not been exploited to constrain the 3D structure of the QGP. Here we use it to subtract the nonflow and recover the full decorrelation map $R_2(\eta_1,\eta_2)$. Since $\beta$ directly controls the ellipticity $\mathcal{E}_2(\eta)$ of the overlap region, the event-wise $\mathcal{E}_2(\eta)$ and $V_2(\eta)$ each split into a component already present for spherical nuclei and a component arising from deformation:
\vspace*{-0.2cm}
\begin{align}\label{eq:3}
\mathcal{E}_2 = \mathcal{E}_{2,\mathrm{sp}}+\mathcal{E}_{2,\beta}\;,\;V_2 = V_{2,\mathrm{sp}}+V_{2,\beta}\;.
\end{align}
These two components are uncorrelated: $\lr{\mathcal{E}_{2,\mathrm{sp}}\mathcal{E}_{2,\beta}^*}=\lr{V_{2,\mathrm{sp}}V_{2,\beta}^*}=0$~\cite{Jia:2021qyu} (Methods~\ref{app:C}). The 2PC flow coefficient receives contributions from both sources plus nonflow contamination, $\delta_{\mathrm{nf}}$ (Fig.~\ref{fig:cartoon}d2):
\vspace*{-0.2cm}
\begin{align}\nonumber
V_{2\Delta} &= V_{2\Delta,\mathrm{sp}}+V_{2\Delta,\beta}+\delta_{\mathrm{nf}}\\\label{eq:4}
          &= a(\eta_1,\eta_2)+ b(\eta_1,\eta_2)\beta^2 + \delta_{\mathrm{nf}}(|\eta_1-\eta_2|)\;.
\end{align}

Each of the two flow terms, $V_{2\Delta,\mathrm{sp}}\!=\!a$ and $V_{2\Delta,\beta}\!=\!b\beta^2$~\cite{Jia:2021qyu,Duguet:2025hwi}, arising respectively from random fluctuations in sampling the nucleon distribution and from dynamical correlations between nucleons, carries its own longitudinal structure (Extended Data~\ref{app:H}). For the deformation-induced term (subscript ``$\beta$'') the decorrelations are accessed by
\vspace*{-0.2cm}
\begin{align}\label{eq:4a}
 R_{2,\beta}(\eta_1,\eta_2)&=\frac{V_{2\Delta,\beta}(\eta_1,\eta_2)}{\sqrt{V_{2\Delta,\beta}(\eta_1,\eta_1)V_{2\Delta,\beta}(\eta_2,\eta_2)}}\\\label{eq:4b} 
r_{2,\beta}(\eta) &= V_{2\Delta,\beta}(\eta,\etaref)/V_{2\Delta,\beta}(-\eta,\etaref)\;.
\end{align}

The longitudinal structure of $V_{2\Delta,\beta}$ can be extracted by comparing isobar pairs $X+X$ and $Y+Y$ of equal mass number, $b=(V_{2\Delta,\mathrm{X}}-V_{2\Delta,\mathrm{Y}})/(\beta_{\mathrm{X}}^2-\beta_{\mathrm{Y}}^2)$.
The subtraction in the numerator eliminates the spherical baseline and the nonflow contributions (Fig.~\ref{fig:cartoon}e), and no value of $\beta$ need be known. For systems with slightly different mass numbers $A$, a multiplicity scale factor is required:
\vspace*{-0.2cm}
\begin{align}\label{eq:5}
V_{2\Delta,\beta}^{\mathrm{X-Y}} = V_{2\Delta,\mathrm{X}}-\frac{\lr{\nch}_{\mathrm{Y}}}{\lr{\nch}_{\mathrm{X}}}V_{2\Delta,\mathrm{Y}}\;.
\end{align}
This works since both the nonflow and the fluctuation-driven baseline in ultra-central collisions scale inversely with the multiplicity $\nch$ (Extended Data~\ref{app:D}).

To validate this framework, we simulate head-on \uuuu{} collisions at $\snn=193$ GeV with and without quadrupole deformation, and \auau{} collisions at $\snn=200$ GeV, using the AMPT transport model (Methods~\ref{app:A} and Table~\ref{tab:1}).

\begin{figure}[htbp]
\centering
\includegraphics[width=1\linewidth]{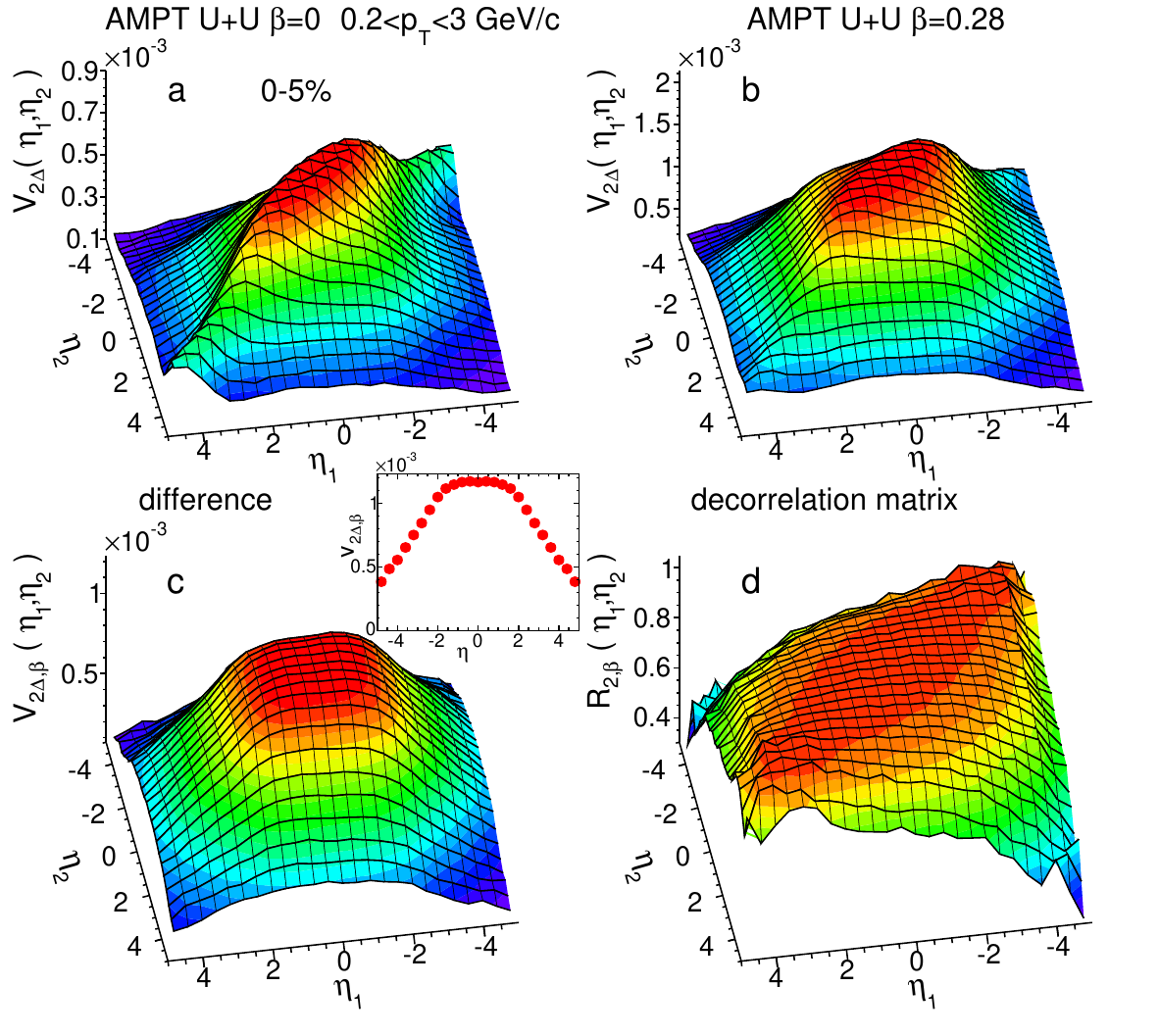}
\caption{\label{fig:v2d} \textbf{Isolating the decorrelation map with a shape difference.} Two-particle elliptic flow $V_{2\Delta}(\eta_1,\eta_2)$ in 0--5\% central U+U collisions: (\textbf{a}) without deformation ($\beta=0$), (\textbf{b}) with deformation ($\beta=0.28$), and (\textbf{c}) their difference $V_{2\Delta,\beta}$, with its diagonal $V_{2\Delta,\beta}(\eta,\eta)$ in the inset, and (\textbf{d}) the decorrelation map $R_{2,\beta}(\eta_1,\eta_2)$ obtained from it via Eq.~\eqref{eq:4a}.}
\end{figure}

{\bf The decorrelation map.} Figure~\ref{fig:v2d}a presents the 2D distribution of $V_{2\Delta}(\eta_1,\eta_2)$ in 0--5\% central U+U collisions without deformation. A prominent nonflow ridge runs along the diagonal~\cite{Pang:2013pma}. Beneath it lies a broader distribution, whose amplitude decreases with the mean rapidity $(\eta_1+\eta_2)/2$ and with $\Deta$, reflecting the decrease of flow with $\eta$ and flow decorrelation, respectively. Nuclear deformation enhances $V_{2\Delta}$ by approximately a factor of two (Fig.~\ref{fig:v2d}b). The difference between the two panels gives $V_{2\Delta,\beta} = b\beta^2$ (Fig.~\ref{fig:v2d}c).

The extracted $V_{2\Delta,\beta}$ is broad. Its diagonal, $V_{2\Delta,\beta}(\eta,\eta)$ (inset), captures only the $\eta$ dependence of the flow magnitude. Dividing it out via Eq.~\eqref{eq:4a} leaves the genuine decorrelation signal $R_{2,\beta}(\eta_1,\eta_2)$ (Fig.~\ref{fig:v2d}d). It depends on the two rapidities separately rather than on their difference alone, so the decorrelations across successive rapidity intervals do not simply add up. The loss is slow up to $|\Deta|\sim2$ and steep beyond. This is compatible with a picture where longitudinal correlations are carried by sources, such as strings, that link only the rapidities they span~\cite{Jia:2024xvl}. This shape, if measured, would therefore constrain both the rapidity extent of these sources and its spread (Extended Data~\ref{app:I}).

\begin{figure}[htbp]
\centering
\includegraphics[width=1\linewidth]{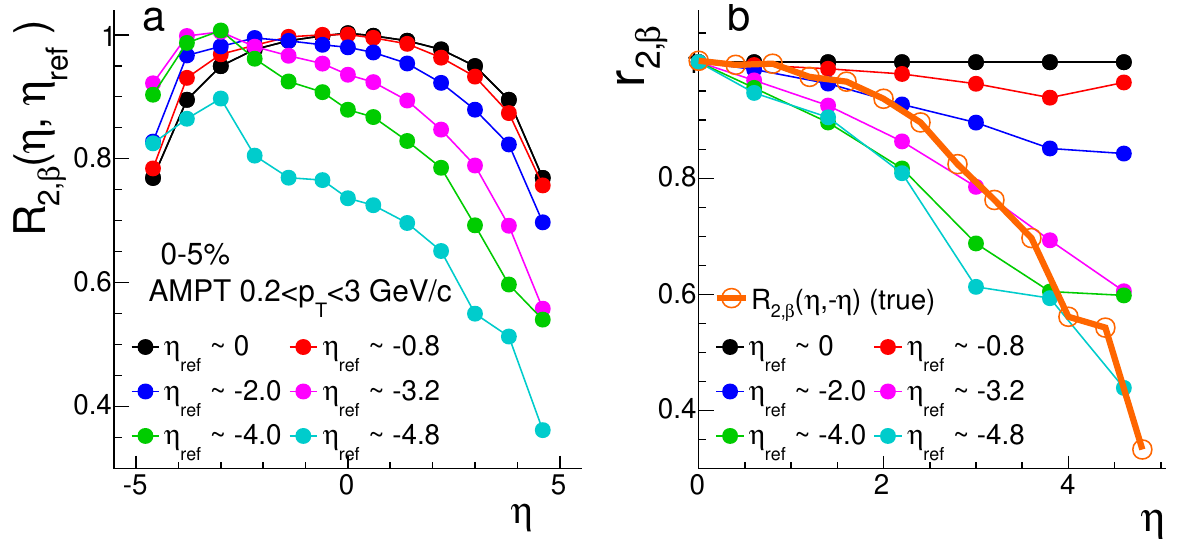}
\caption{\label{fig:proj} \textbf{$r_2(\eta)$ is not the decorrelation it was proposed to measure.} (\textbf{a}) $R_{2,\beta}(\eta,\etaref)$ as a function of $\eta$ in narrow slices of $\etaref$. Values further below unity indicate stronger decorrelation. (\textbf{b}) $r_{2,\beta}$ calculated via Eq.~\eqref{eq:4b} (filled symbols), compared with the true decorrelation between the mirror rapidities, $R_{2,\beta}(\eta,-\eta)$ (open symbols).}
\end{figure}

Figure~\ref{fig:proj}a shows $R_{2,\beta}(\eta,\etaref)$ as a function of $\eta$ for narrow slices of $\etaref$, which quantifies how the flow at $\eta$ decorrelates from a reference at $\etaref$. The further $R_{2,\beta}$ falls below unity, the stronger the decorrelation. The pattern depends strongly on choice of $\etaref$. For a far-backward reference, $\etaref\sim -5$, $R_{2,\beta}$ falls steadily over almost the entire $\eta$ range. For a mid-rapidity reference, $\etaref\sim0$, it stays close to unity out to $\eta\sim 2$ and only then drops quickly.

This reference dependence is either hidden or distorted by $r_2$. Figure~\ref{fig:proj}b shows $r_{2,\beta}(\eta)$ obtained from the same $R_{2,\beta}$, together with the true decorrelation between $\eta$ and $-\eta$, $R_{2,\beta}(\eta,-\eta)$. Its values change substantially with $\etaref$, so the common reading of $r_2(\eta)$ as the true signal $R_{2,\beta}(\eta,-\eta)$~\cite{CMS:2015xmx,ATLAS:2020sgl,Behera:2020mol} holds only at the single point $\eta=|\etaref|$. In fact, over the range $|\eta|<|\etaref|$ used in practice, $r_2$ falls below the true value and so overstates the decorrelation, while it understates the decorrelation for $|\eta|>|\etaref|$. Extracting physical information from $r_2$ thus requires assumptions, such as independence from $\etaref$, which is invalid, and symmetric collision systems (Extended Data~\ref{app:I}).

The decorrelations above are obtained from the ideal difference between U+U simulations with and without deformation. In practice only Au+Au and U+U collisions are available, and their mass numbers differ by 20\%. The multiplicity-scaled subtraction of Eq.~\eqref{eq:5} absorbs this difference and removes the nonflow, such that the resulting $R_{2,\beta}(\eta_1,\eta_2)$ and its projections reproduce the ideal case (Extended Data Fig.~\ref{fig:uau}).

{\bf One profile for both components.} We now show that the spherical baseline $V_{2\Delta,\mathrm{sp}}(\eta_1,\eta_2)$ has a longitudinal structure remarkably similar to that of $V_{2\Delta,\beta}(\eta_1,\eta_2)$. The longitudinal structure of $V_2$ is governed by the forward- and backward-going eccentricities together. The initial geometry at forward (backward) rapidity aligns more closely with $\mathcal{E}_{2,\mathrm{F}}$ ($\mathcal{E}_{2,\mathrm{B}}$), while at mid-rapidity the flow responds to the average $\mathcal{E}_{2}\sim (\mathcal{E}_{2,\mathrm{F}}+\mathcal{E}_{2,\mathrm{B}})/2$~\cite{Jia:2014ysa,Behera:2020mol} (Fig.~\ref{fig:cartoon}c).

Projecting $V_2(\eta)$ onto $\mathcal{E}_{2,\mathrm{B}}$ and onto its spherical-baseline and deformation parts, $\mathcal{E}_{2,\mathrm{B,sp}}$ and $\mathcal{E}_{2,\mathrm{B},\beta}$, gives three flow coefficients, $v_{2,\varepsilon_{\mathrm{B}}}(\eta)$, $v_{2,\varepsilon_{\mathrm{B,sp}}}(\eta)$ and $v_{2,\varepsilon_{\mathrm{B},\beta}}(\eta)$ (Methods~\ref{app:C}). Being correlations with an initial-state quantity, they are free of nonflow and show directly how the flow at $\eta$ decorrelates from a fixed reference geometry. 

\begin{figure}[htbp]
\centering
\includegraphics[width=1\linewidth]{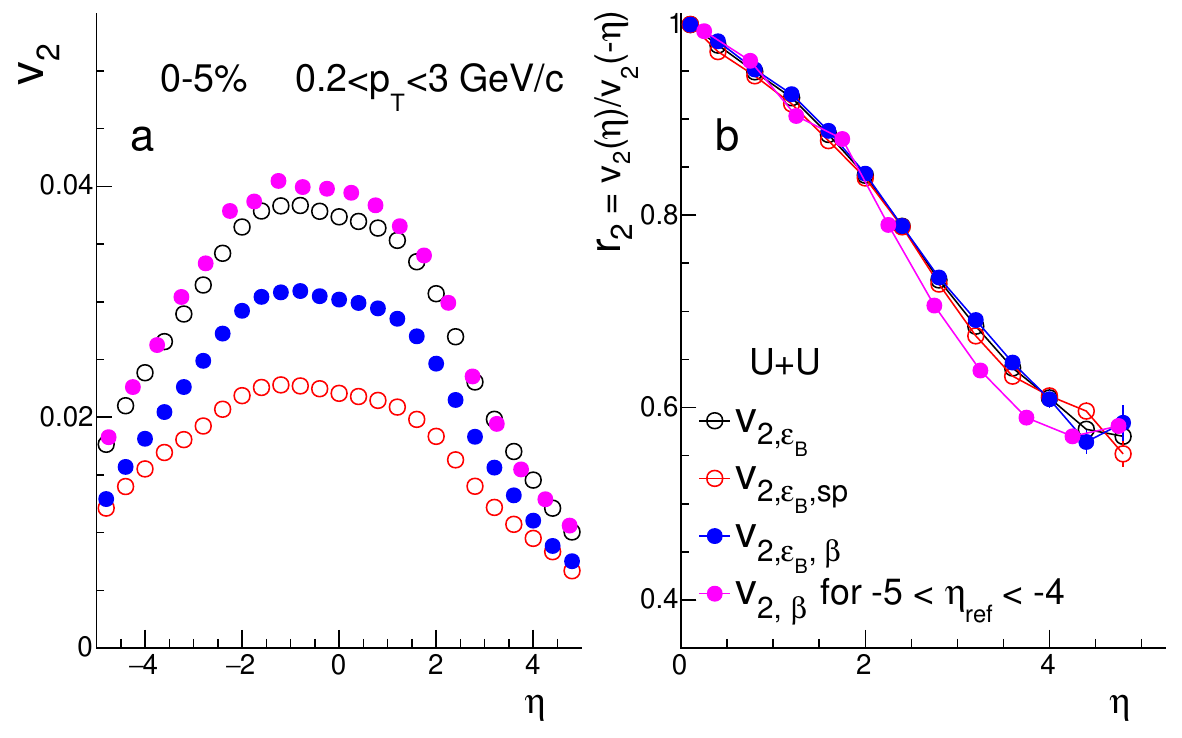}
\caption{\label{fig:univ} \textbf{Universality of longitudinal decorrelations.} (\textbf{a}) Comparison of flow calculated using backward-going eccentricity: $v_{2,\varepsilon_{\mathrm{B}}}$ (open black), $v_{2,\varepsilon_{\mathrm{B,sp}}}$ (open red), $v_{2,\varepsilon_{\mathrm{B},\beta}}$ (solid blue) defined in Methods Eq.~\eqref{eq:7}, and deformation-driven flow $v_{2,\beta}$ (solid pink) from Methods Eq.~\eqref{eq:8}. (\textbf{b}) Ratios of the values at $\eta$ and $-\eta$ for the four distributions in panel (a).}
\end{figure}

All three are strongly asymmetric about $\eta=0$ due to decorrelations (Fig.~\ref{fig:univ}a). We overlay these with the deformation-induced flow $v_{2,\beta}(\eta)$, calculated from $V_{2\Delta,\beta}(\eta,\etaref)$ of Fig.~\ref{fig:v2d}c as $v_{2,\beta}(\eta)=V_{2\Delta,\beta}(\eta,\etaref)/\sqrt{V_{2\Delta,\beta}(\etaref,\etaref)}$ using reference particles at $-5<\etaref<-4$. Flow at backward rapidity is itself driven mainly by $\mathcal{E}_{2,\mathrm{B}}$, so the shape of $v_{2,\beta}(\eta)$ derived from this reference choice should track that of $v_{2,\varepsilon_{\mathrm{B},\beta}}(\eta)$.

By analogy with Eq.~\eqref{eq:2}, we quantify the decorrelations of the distributions in Fig.~\ref{fig:univ}a by their forward-to-backward ratios $r_{2,X}(\eta)=v_{2,X}(\eta)/v_{2,X}(-\eta)$ with $X\in\{\varepsilon_{\mathrm{B}},\,\varepsilon_{\mathrm{B,sp}},\,\varepsilon_{\mathrm{B},\beta},\,\beta\}$. For $X=\beta$ this is equivalent to Eq.~\eqref{eq:4b}. All four ratios decrease strongly with $\eta$, and by remarkably similar amounts (Fig.~\ref{fig:univ}b), indicating that the flow components corresponding to spherical nuclei and to nuclear deformation possess nearly identical longitudinal structure~\footnote{Small differences are observed in 0--2\% central collisions; see Extended Data~\ref{app:F}--\ref{app:G} for detailed discussion.}. Nuclear deformation therefore changes only the magnitude of $V_{2\Delta}(\eta_1,\eta_2)$, not its longitudinal profile. This universality is physically expected: both components arise from the same participating nucleons, whether those nucleons are arranged in a deformed or spherical configuration. It implies that the longitudinal structure of $V_{2\Delta}(\eta_1,\eta_2)$ can be recovered from $V_{2\Delta,\beta}(\eta_1,\eta_2)$ over the full pseudorapidity range, free from nonflow contamination.

{\bf The bias in flow measurements.} Beyond imaging the initial state, the extracted $R_2(\eta_1,\eta_2)$ has direct consequences for the flow measurements themselves. The $\pT$-differential flow of particles of interest (POI) at rapidity $\eta$ is usually measured by correlating them with reference particles at forward and backward rapidities $\pm\etaref$ (the three-subevent method)~\cite{Poskanzer:1998yz}:
\vspace*{-0.2cm}
\small{\begin{align}\nonumber
v_2(\pT)_{\mathrm{est.}}&=\!\sqrt{\!\frac{\lr{V_2(\pT,\eta) V_2^*(\etaref)}\lr{V_2(\pT,\eta) V_2^*(-\etaref)}}{\lr{V_2(\etaref) V_2^*(-\etaref)}}} \\\nonumber
&=\! v_2(\pT,\eta) B_2\\\label{eq:bias}
B_2(\eta,\etaref)&=\sqrt{\frac{R_2(\eta,\etaref)\,R_2(\eta,-\etaref)}{R_2(\etaref,-\etaref)}}\;.
\end{align}}\normalsize
The measured flow thus differs from the true one by a bias factor $B_2$ that depends on $\eta$ and $\etaref$. In Eq.~\eqref{eq:bias} the decorrelations cancel only under one condition: $R_2(\eta,\etaref)\,R_2(\eta,-\etaref)$ should equal $R_2(\etaref,-\etaref)$.  This is satisfied whenever the losses of correlation across successive rapidity intervals simply add up, and only within $|\eta|<|\etaref|$. The exponential form $R_2=e^{-F_2|\Deta|}$ ($\approx 1-F_2|\Deta|$) often assumed for $r_2$ is one such case~\cite{CMS:2015xmx,ATLAS:2017rij,ATLAS:2020sgl,Nie:2020trj,STAR:2025vmb} (Extended Data~\ref{app:I}). But in practice, more correlation is lost across the full gap than across its two halves combined, $R_2(\etaref,-\etaref)< R_2(0,\etaref)\,R_2(0,-\etaref)$, so $B_2$ exceeds unity.

\begin{figure}[htbp]
\centering
\includegraphics[width=0.8\linewidth]{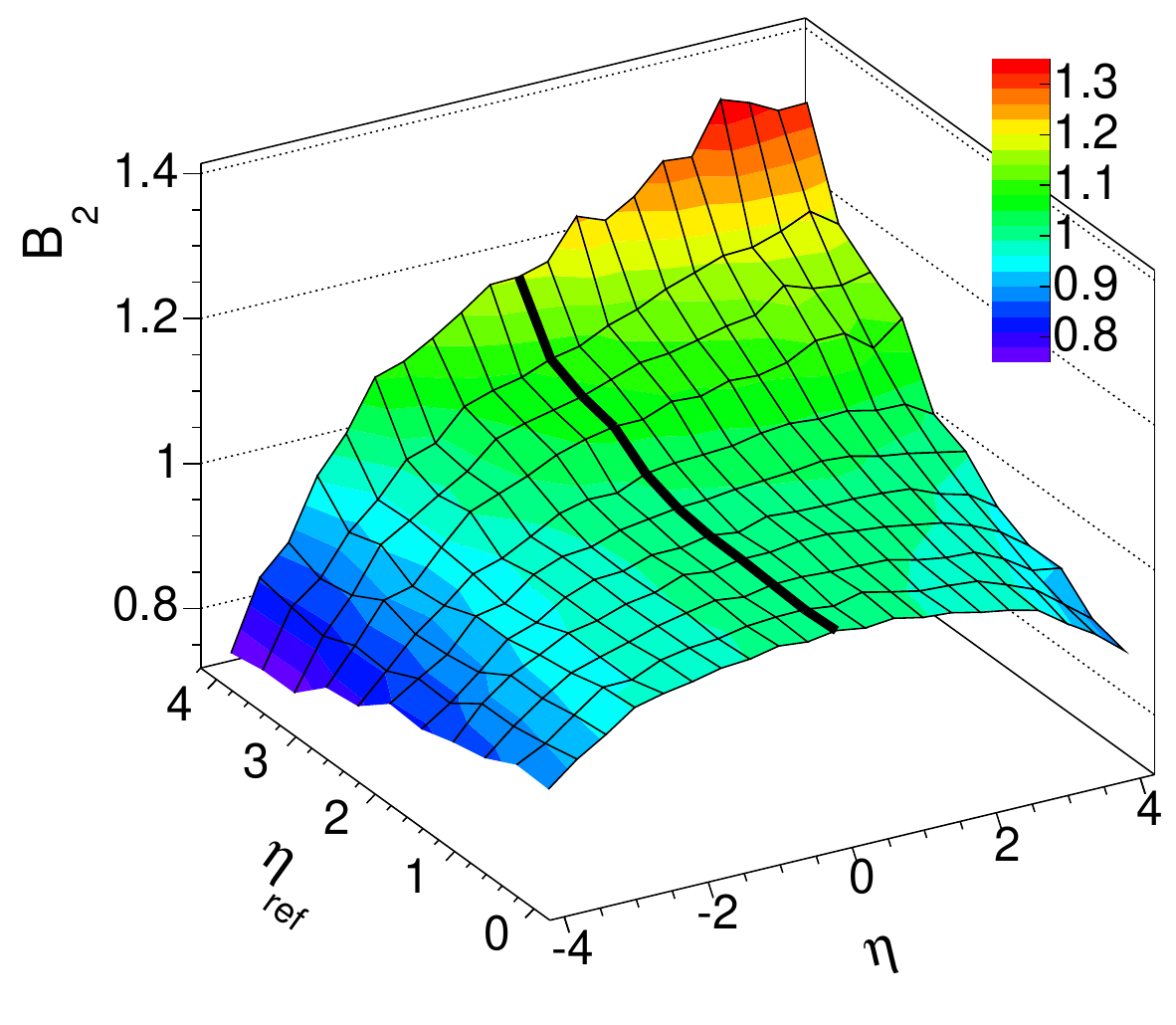}
\caption{\label{fig:dec} \textbf{How decorrelations bias flow measurements.} The bias factor $B_2(\eta,\etaref)$ of Eq.~\eqref{eq:bias} for $v_2(\pT)$ measured via the subevent method, as a function of the POI rapidity $\eta$ and the reference rapidity $\etaref$, calculated from $R_{2,\beta}(\eta_1,\eta_2)$ in Fig.~\ref{fig:v2d}d.  $B_2$ is an even function of $\etaref$ by definition in Eq.~\eqref{eq:bias}, so only the $\etaref>0$ region is shown. The thick black line indicates the bias factor for a POI at $\eta=0$.}
\end{figure}

The universality established in Fig.~\ref{fig:univ} allows $B_2$ to be evaluated from $R_{2,\beta}(\eta_1,\eta_2)$ of Fig.~\ref{fig:v2d}d. The results are shown in Fig.~\ref{fig:dec}. The bias factor exceeds unity for POI at mid-rapidity ($\eta\sim0$), by $\sim15$\% at $\etaref\approx4$ and $\sim8$\% for the typical RHIC choice $|\etaref|\approx3$. A bias factor of this kind exists for every three-subevent measurement.  It differs from one experiment to another and is present in any $v_n(\pT)$ measurement that uses separated reference rapidities. 

Such biases are expected to be more severe in small systems, where the decorrelations are stronger. In addition, the flow signal there is weak and the nonflow is large. Experiments therefore combine additional subevents to suppress it, and each such combination carries a decorrelation bias of its own that must be evaluated separately. For $v_3$, whose decorrelations are much stronger than those of $v_2$, the bias should be considerably larger. This $v_3$ bias bears directly on the small-system puzzle at RHIC, where experiments with very different rapidity acceptances report conflicting $v_3$ orderings~\cite{PHENIX:2018lia,STAR:2023wmd}, and longitudinal decorrelations could be an important source of this discrepancy~\cite{Zhao:2022ugy} (Extended Data~\ref{app:K}). Quantifying these biases requires the nonflow-free $R_n$ maps, which are needed both to constrain the initial state and to measure the flow itself reliably.
 
{\bf Summary.} We compare isobar-like systems whose ellipticity is strongly affected by nuclear deformation while the nonflow stays nearly identical. We find that deformation-induced and fluctuation-driven flow decorrelate identically, so a shape difference provides access to the full decorrelation, as long as the difference is large enough to measure. This comparison allows us to recover the complete longitudinal structure of $V_{2\Delta}(\eta_1,\eta_2)$ and the factorization ratio $R_2(\eta_1,\eta_2)$, free of the bias of the traditional decorrelation observable $r_2(\eta)$. The structure of $R_2(\eta_1,\eta_2)$ shows that the losses of correlation across successive rapidity intervals do not simply add up, resulting in significant bias even in ordinary flow measurements. The same property makes $r_2(\eta)$ overestimate the decorrelation it was proposed to measure. Neither bias has been quantified before, yet both must be known before results obtained with different rapidity combinations can be compared.

{\bf Outlook.} The maps obtained here are predictions for what the data should show once the nonflow is removed, and they open avenues on both sides. For theory, the same construction applies to any description of the initial state and of the subsequent evolution, so comparing our predicted maps with measured ones would constrain how energy is deposited along rapidity, how those sources fluctuate, and how that evolution reshapes them~\cite{Jia:2024xvl}. For experiment, the map allows us to evaluate the bias of any chosen acceptance, and with it the role of decorrelation in the small-system $v_3$ puzzle~\cite{PHENIX:2018lia,STAR:2023wmd}. Our approach matters most where the rapidity coverage is limited, and in the small systems, where the bias is largest. The system pairs required have already been recorded: $^{238}$U+$^{238}$U/$^{197}$Au+$^{197}$Au, $^{96}$Ru+$^{96}$Ru/$^{96}$Zr+$^{96}$Zr and $d$+Au/$^{16}$O+$^{16}$O at RHIC, and $^{16}$O+$^{16}$O/$^{20}$Ne+$^{20}$Ne at the LHC, spanning small to large collision systems.

The same strategy extends beyond elliptic flow. Any pair of similar-mass systems differing in a chosen structural property gives access to a different longitudinal geometry mode. The $^{96}$Ru and $^{96}$Zr pair, for example, differs in both octupole deformation and radial profile~\cite{Jia:2021oyt,Xu:2021uar}, and therefore opens the longitudinal structure of the triangularity $\mathcal{E}_3(\eta)$ and of the fireball size $S(\eta)$, respectively (Fig.~\ref{fig:cartoon}b). Together these would map all the leading longitudinal geometry modes of the QGP.


\section*{Methods}

\subsection{Model setup and simulation}\label{app:A}
We study \uuuu{} collisions at $\snn=193$ GeV and \auau{} collisions at $\snn=200$ GeV, the same energies chosen by experiments~\cite{STAR:2024eky}. Uranium exhibits large prolate deformation, while gold is nearly spherical with mild oblate deformation (Table~\ref{tab:1}). We focus on the ultra-central collision (UCC) region where deformation effects are most pronounced. 

We use the AMPT transport model~\cite{Lin:2004en}, which simulates the full 3D evolution of heavy-ion collisions. The nucleon distribution is sampled from a Woods-Saxon parameterization in polar coordinates: $\rho(r,\theta) \propto \left\{1+\exp\left[\left(r-R_0\left(1+\beta Y_{2,0}(\theta)\right)\right)/a_0\right]\right\}^{-1}$. We simulate the collisions in the string-melting mode with a partonic cross-section of 3~mb for the three configurations listed in Table~\ref{tab:1}. We calculate eccentricity separately for forward-going, backward-going, and all participating nucleons, denoted as $\mathcal{E}_{2,\mathrm{F}}$, $\mathcal{E}_{2,\mathrm{B}}$, and $\mathcal{E}_{2}$, respectively. Here ``forward-going'' means the participating nucleons of the nucleus travelling in the $+z$ direction, and ``backward-going'' those of the other nucleus. Particles for 2PC analysis are selected from the transverse momentum range $0.2<\pT<3$ GeV/$c$. Event centrality is defined by $\nch$ in $|\eta|<0.5$, and the multiplicity ratio in Eq.~\eqref{eq:5} is taken over the same window.  Correlations are accumulated in bins of width $\delta\eta=0.4$ over $|\eta|<5$. 

The observables $R_2(\eta_1,\eta_2)$ and $r_2(\eta)$ are constructed as ratios from the final-state flow. Owing to the relation $V_2(\eta) \propto \mathcal{E}_2(\eta)$, they are largely insensitive to model dependence in the final state~\cite{Pang:2015zrq,Wu:2018cpc} and thus mainly reflect the longitudinal structure of the initial state. AMPT reproduces the measured rapidity distribution of particle production at RHIC~\cite{Lin:2004en}, the measured $r_n(\eta)$ when used as a hydrodynamic initial condition~\cite{Pang:2015zrq,Wu:2018cpc}, and, in the string-melting configuration used here, the isobar ratios of $v_2$, $v_3$ and multiplicity~\cite{Jia:2021oyt,Jia:2022qgl}. Our predictions should therefore provide quantitative expectations for the rapidity structure of elliptic flow, which can be validated in future real data analysis. 

\begin{table}[!h]
\centering
\caption{\label{tab:1} Nuclei species and their quadrupole deformation $\beta$, radius $R_0$ and skin thickness $a_0$ used in AMPT simulation.} 
\small{\begin{tabular}{c|c|c|c}\hline
  System        &$^{238}$U+$^{238}$U  & $^{238}$U+$^{238}$U & $^{197}$Au+$^{197}$Au \\\hline
 $\beta$,$R_0$(fm),$a_0$(fm)\!        & 0.28, 6.81, 0.55     & 0, 6.81, 0.55         & -0.14, 6.62, 0.52  \\\hline
\end{tabular}}\normalsize
\end{table}

\subsection{Notation conventions}\label{app:B}
The subscripts used in this work have the following meanings:
\begin{itemize}[noitemsep]
\item ``sp'' = spherical baseline, i.e., $\beta=0$
\item ``$\beta$'' = deformation-induced contributions  
\item ``$\varepsilon_{\mathrm{B}}$'' = projection onto backward eccentricity $\mathcal{E}_{2,\mathrm{B}}$
\item Combined subscripts e.g., ``$\varepsilon_{\mathrm{B},\beta}$'' = deformation-induced flow projected onto backward eccentricity
\end{itemize}

\subsection{Flow projected onto the initial eccentricity}\label{app:C}
Analogous to Eq.~\eqref{eq:3}, the backward-going eccentricity contains independent spherical-baseline and deformation contributions, $\mathcal{E}_{2,\mathrm{B}} = \mathcal{E}_{2,\mathrm{B,sp}}+\mathcal{E}_{2,\mathrm{B},\beta}$.
We access the longitudinal structure of $V_2$ by projecting it onto $\mathcal{E}_{2,\mathrm{B}}$ and onto these two components, which defines the three observables of Fig.~\ref{fig:univ}:
\small{\begin{align}\nonumber
&v_{2,\varepsilon_{\mathrm{B}}}(\eta) \equiv\frac{\lr{V_2(\eta) \mathcal{E}_{2,\mathrm{B}}^*}}{\sqrt{\lr{\mathcal{E}_{2,\mathrm{B}}\mathcal{E}_{2,\mathrm{B}}^*}}}\;,\\\label{eq:7}
&v_{2,\varepsilon_{\mathrm{B,sp}}}(\eta) \equiv \frac{\lr{V_2(\eta) \mathcal{E}_{2,\mathrm{B,sp}}^*}}{\sqrt{\lr{\mathcal{E}_{2,\mathrm{B,sp}}\mathcal{E}_{2,\mathrm{B,sp}}^*}}} \simeq \frac{\lr{V_2(\eta) \mathcal{E}_{2,\mathrm{B}}^*}_{\beta=0}}{\sqrt{\lr{\mathcal{E}_{2,\mathrm{B}}\mathcal{E}_{2,\mathrm{B}}^*}_{\beta=0}}}\;,\\\nonumber
&v_{2,\varepsilon_{\mathrm{B},\beta}} (\eta)\! \equiv\! \frac{\lr{V_2 (\eta)\mathcal{E}_{2,\mathrm{B},\beta}^*}}{\sqrt{\lr{\mathcal{E}_{2,\mathrm{B},\beta}\mathcal{E}_{2,\mathrm{B},\beta}^*}}}\!\simeq\!\frac{\lr{V_2 (\eta)\mathcal{E}_{2,\mathrm{B}}^*}_{\beta\!=0\!.28}\!-\!\lr{V_2 (\eta)\mathcal{E}_{2,\mathrm{B}}^*}_{\beta\!=\!0}\!}{\sqrt{\lr{\mathcal{E}_{2,\mathrm{B}}\mathcal{E}_{2,\mathrm{B}}^*}_{\beta\!=\!0.28}\!-\!\lr{\mathcal{E}_{2,\mathrm{B}}\mathcal{E}_{2,\mathrm{B}}^*}_{\beta\!=\!0}}\!}\;.
\end{align}}\normalsize
The last two relations express the spherical and deformation projections using quantities computable from the $\beta=0.28$ and $\beta=0$ simulation samples. 

The deformation-induced flow overlaid in Fig.~\ref{fig:univ}a is obtained instead from the two-particle correlator, with reference particles at $-5<\etaref<-4$:
\begin{align}\label{eq:8}
v_{2,\beta} (\eta)= \frac{V_{2\Delta,\beta}(\eta,\etaref)}{\sqrt{V_{2\Delta,\beta}(\etaref,\etaref)}}\;.
\end{align}

In this decomposition, the deformation-induced eccentricity $\mathcal{E}_{2,\mathrm{B},\beta}$ and the spherical-baseline eccentricity $\mathcal{E}_{2,\mathrm{B,sp}}$ are statistically independent, $\langle \mathcal{E}_{2,\mathrm{B,sp}}  \mathcal{E}_{2,\mathrm{B},\beta}^* \rangle = 0$, so the cross term averages to zero and we get,
\begin{align*}
\lr{\varepsilon_{2,\mathrm{B}}^2} &= \lr{\varepsilon_{2,\mathrm{B,sp}}^2} + \lr{\varepsilon_{2,\mathrm{B},\beta}^2}\;,\\
\lr{v_{2,\varepsilon_{\mathrm{B}}}^2} &= \lr{v_{2,\varepsilon_{\mathrm{B,sp}}}^2} + \lr{v_{2,\varepsilon_{\mathrm{B},\beta}}^2}\;,
\end{align*}
the second following from the linear hydrodynamic response $v_2 \approx \kappa \varepsilon_2$.

\renewcommand{\figurename}{\textbf{Extended Data Fig.}}
\section*{Extended Data}

\subsection{Why the multiplicity scaling works}\label{app:D}
The multiplicity-scaled subtraction in Eq.~\eqref{eq:5} works because both unwanted terms scale inversely with the system size. 2PC nonflow is diluted as $\delta_{\mathrm{nf}}\propto 1/\nch$, and in ultra-central collisions $\nch\propto N_{\mathrm{part}}\approx 2A$, so $\delta_{\mathrm{nf}}\propto 1/A$. The same scaling underlies experimental nonflow subtractions~\cite{CMS:2013jlh,STAR:2022pfn}. The fluctuation-driven baseline follows the same scaling in ultra-central collisions, $V_{2\Delta,\mathrm{sp}}\propto 1/A$~\cite{Blaizot:2014wba}. Hence $V_{2\Delta,\beta}^{\mathrm{X-Y}}$ in UCC captures the structure of $b(\eta_1,\eta_2)$, free of nonflow and spherical-baseline contributions, as Extended Data Fig.~\ref{fig:uau} demonstrates.

\subsection{Nonflow removal in the U+U versus Au+Au comparison}\label{app:E}

\begin{figure}[htbp]
\vspace*{0.2cm}
\centering
\includegraphics[width=0.99\linewidth]{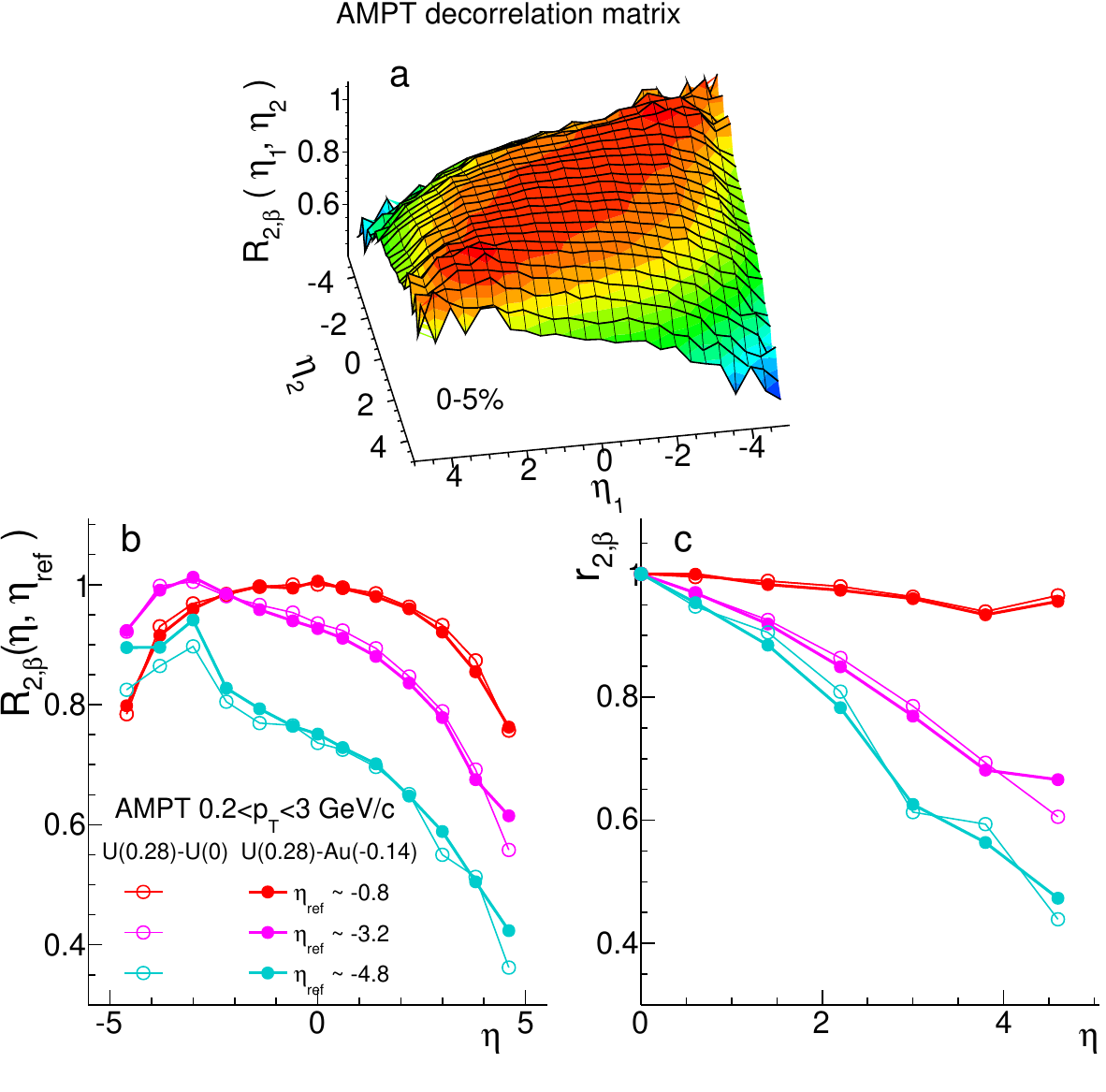}
\caption{\label{fig:uau} \textbf{Nonflow removal using isobar-like systems.} (\textbf{a}) $R_{2,\beta}(\eta_1,\eta_2)$ obtained from $V_{2\Delta,\beta}^{\mathrm{U-Au}}$ via Eq.~\eqref{eq:5}, with nearly the same structure as the ideal case in Fig.~\ref{fig:v2d}d. (\textbf{b}) $R_{2,\beta}(\eta,\etaref)$ as a function of $\eta$ in slices of $\etaref$, compared with Fig.~\ref{fig:proj}a. (\textbf{c}) $r_{2,\beta}$ calculated via Eq.~\eqref{eq:4b}, compared with Fig.~\ref{fig:proj}b. The agreement demonstrates the validity of the nonflow subtraction.}
\end{figure}

Figure~\ref{fig:uau}a presents $R_{2,\beta}(\eta_1,\eta_2)$ derived from $V_{2\Delta,\beta}^{\mathrm{U-Au}}$. Its shape closely matches the ideal case of Fig.~\ref{fig:v2d}d, and the projections (Figs.~\ref{fig:uau}b,c) reproduce those of Figs.~\ref{fig:proj}a,b. The system comparison thus removes the nonflow in the realizable U+U/Au+Au comparison, not just in the idealized $\beta$-switching test.

\subsection{Decorrelations in ultra-central 0--2\% collisions}\label{app:F}

\begin{figure}[htbp]
\centering
\includegraphics[width=1\linewidth]{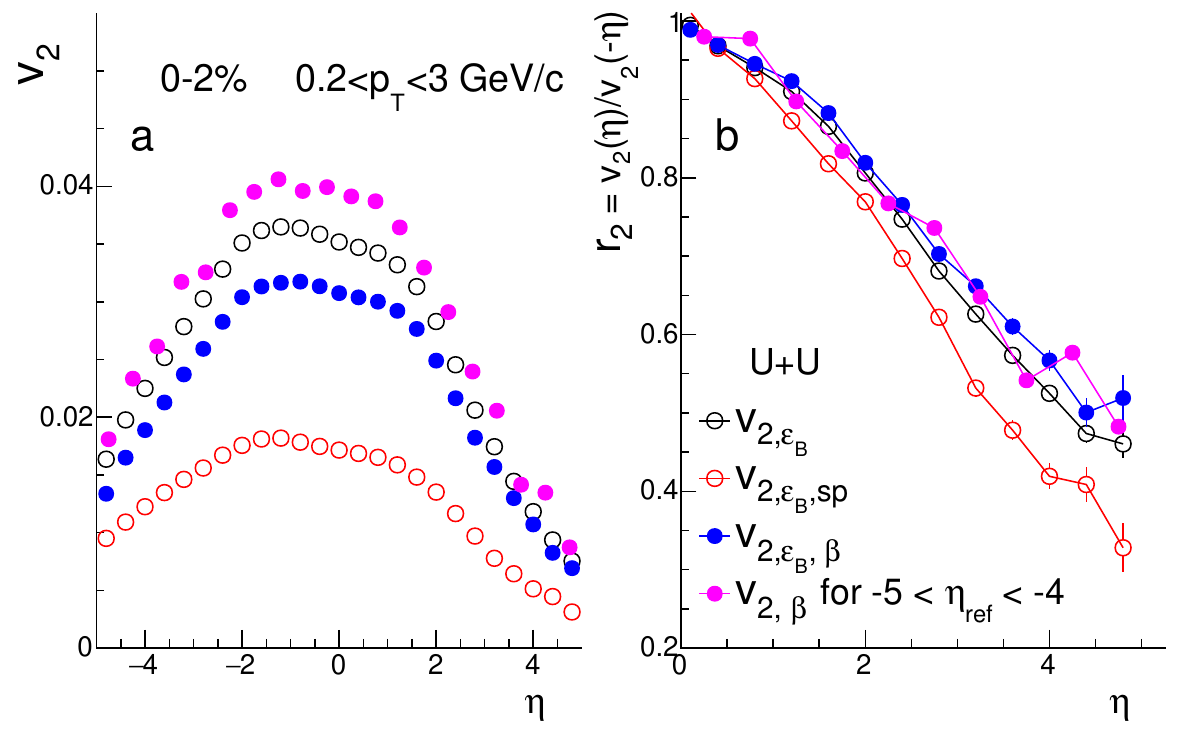}
\caption{\label{fig:ed1} \textbf{Decorrelations in ultra-central collisions.} Same as Fig.~\ref{fig:univ} but for 0--2\% most central U+U collisions. (\textbf{a}) Comparison of flow calculated using backward-going eccentricity: $v_{2,\varepsilon_{\mathrm{B}}}$ (open black), $v_{2,\varepsilon_{\mathrm{B,sp}}}$ (open red), $v_{2,\varepsilon_{\mathrm{B},\beta}}$ (solid blue), and $v_{2,\beta}$ (solid pink). (\textbf{b}) Corresponding ratios. Subtle differences appear in this extreme centrality range, due to alignment of the nuclear axes.}
\end{figure}

Figure~\ref{fig:ed1} displays longitudinal decorrelations in 0--2\% most central collisions, analogous to Fig.~\ref{fig:univ}. While the general trends resemble those observed in 0--5\% centrality, the overall decorrelation strength is noticeably stronger, consistent with previous findings on the centrality dependence of flow decorrelations~\cite{CMS:2015xmx,ATLAS:2017rij}. Interestingly, $r_{2,\varepsilon_{\mathrm{B,sp}}}$ (open red) shows a slightly stronger decrease compared to $r_{2,\varepsilon_{\mathrm{B}}}$ (open black), implying that decorrelations of $V_{2\Delta,\mathrm{sp}}$ are somewhat stronger than those of $V_{2\Delta,\beta}$. However, the maximum difference between these ratios, at $\eta\approx4.8$, is about 0.08, small compared with the total decorrelation of $r_2$, which is $1-r_2 = 0.5$. This subtle difference between the deformation-induced component and its spherical baseline appears only in extremely central collisions. It comes from the tendency of the symmetry axes of the two colliding nuclei to align in ultra-central collisions, due to the requirement of large $\nch$, as discussed below.

\subsection{Alignment of forward- and backward-going eccentricities}\label{app:G}
\begin{figure}[!h]
\includegraphics[width=0.75\linewidth]{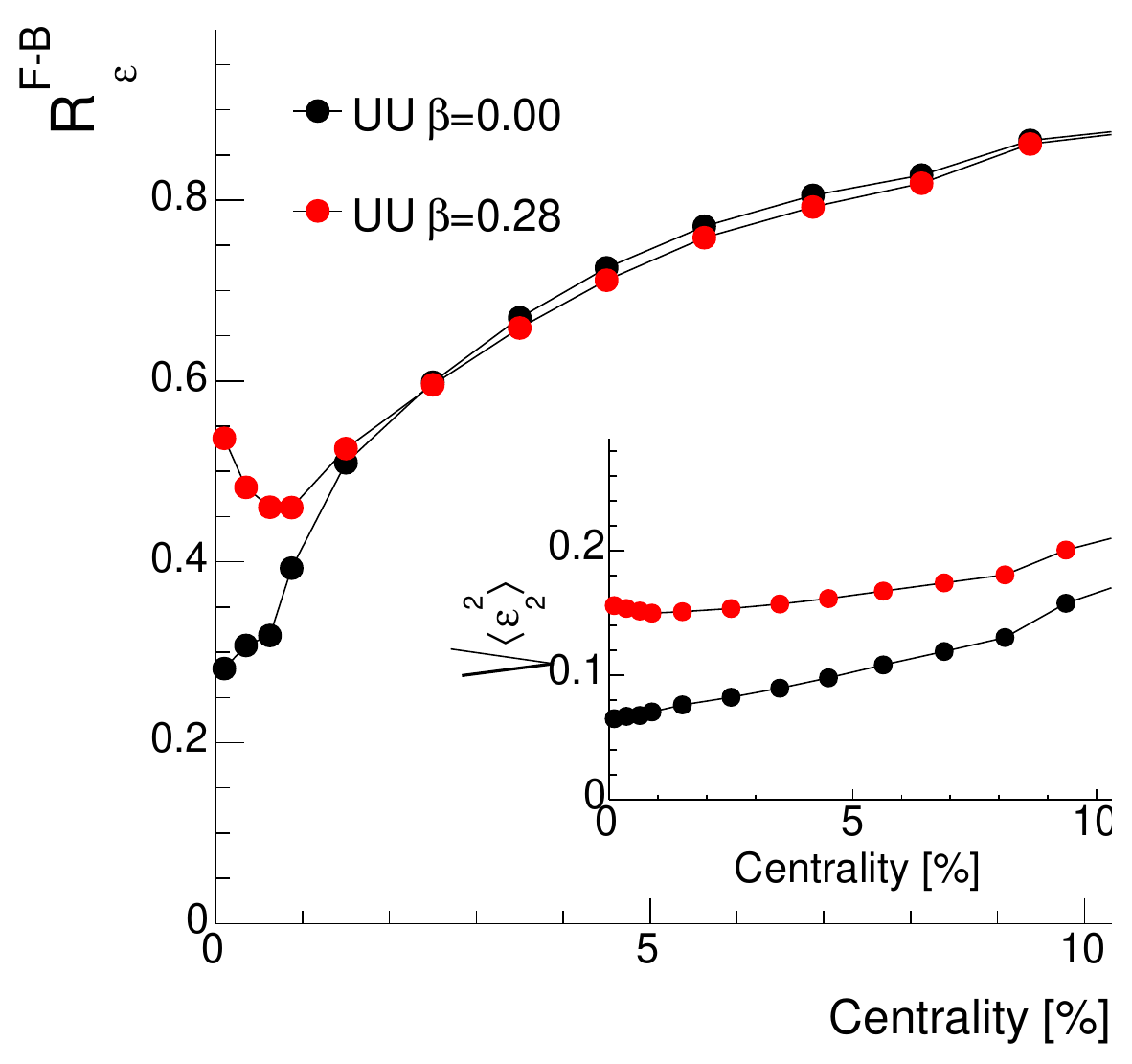}
\caption{\label{fig:ed2} \textbf{Alignment of forward- and backward-going eccentricities.} Centrality dependence of $R_{\varepsilon}^{\mathrm{F-B}}$ defined in Eq.~\eqref{eq:a0} for U+U collisions with and without deformation. The inset shows the root-mean-square values of $\varepsilon_2$ for the two cases. Deformation always enhances flow across a wide centrality range, even where $\mathcal{E}_{2,\mathrm{B}}$ and $\mathcal{E}_{2,\mathrm{F}}$ are not aligned.}
\end{figure}

The longitudinal structure of QGP's initial ellipticity interpolates between $\mathcal{E}_{2,\mathrm{F}}$ at forward rapidity and $\mathcal{E}_{2,\mathrm{B}}$ at backward rapidity. Therefore, the decorrelations in Figs.~\ref{fig:univ} and \ref{fig:ed1} should reflect the decorrelations between $\mathcal{E}_{2,\mathrm{F}}$ and $\mathcal{E}_{2,\mathrm{B}}$, characterized by:
\begin{align}\label{eq:a0}
R_{\varepsilon}^{\mathrm{F-B}}  = \frac{\lr{\mathcal{E}_{2,\mathrm{F}}\mathcal{E}_{2,\mathrm{B}}^*}}{\sqrt{\lr{\mathcal{E}_{2,\mathrm{F}}\mathcal{E}_{2,\mathrm{F}}^*}\lr{\mathcal{E}_{2,\mathrm{B}}\mathcal{E}_{2,\mathrm{B}}^*}}}\;.
\end{align}
The centrality dependence of $R_{\varepsilon}^{\mathrm{F-B}}$ is displayed in Fig.~\ref{fig:ed2} for U+U collisions with and without deformation. The presence of deformation increases the values of $R_{\varepsilon}^{\mathrm{F-B}}$ in the 0--2\% centrality range, implying weaker decorrelations (stronger alignment). Beyond 2\% centrality, deformation has negligible impact on $R_{\varepsilon}^{\mathrm{F-B}}$. These behaviours naturally explain the ordering of decorrelations observed in Fig.~\ref{fig:ed1}b. 

In contrast, the eccentricity magnitude, quantified by mean-square values, is consistently enhanced by deformation over a much broader centrality range, as shown in the inset. Evidently, deformation always enhances $\lr{\varepsilon_2^2}$ but does not necessarily lead to better alignment of the two eccentricity vectors $\mathcal{E}_{2,\mathrm{F}}$ and $\mathcal{E}_{2,\mathrm{B}}$~\footnote{Both $\mathcal{E}_{2,\mathrm{F}}$ and $\mathcal{E}_{2,\mathrm{B}}$ contain a deformation component $\mathcal{E}_{2,\beta}$ that is statistically uncorrelated between the two nuclei. The mean square value of the total eccentricity $\mathcal{E}_{2}=(\mathcal{E}_{2,\mathrm{F}}+\mathcal{E}_{2,\mathrm{B}})/2$ is then enhanced by $\lr{\mathcal{E}_{2,\beta}\mathcal{E}_{2,\beta}^*}/2$. For full alignment of the nuclear axes, the enhancement would be $\lr{\mathcal{E}_{2,\beta}\mathcal{E}_{2,\beta}^*}$.}. The enhanced alignment of eccentricity vectors observed in the 0--2\% centrality range results from the requirement of large $\nch$, which favours alignment of the nuclear symmetry axes. The selection of central but not ultra-central collisions does not fully enforce such alignment, and the orientations of the two colliding nuclei therefore remain largely random with respect to each other.

\subsection{Nuclear shape and eccentricity}\label{app:H}
The decomposition of Eq.~\eqref{eq:4} is usually read as a spherical baseline plus a deformation correction, but it has a more general origin. In a perturbative treatment of initial-state fluctuations, the mean squared eccentricity of the overlap region separates into a random and a dynamical term~\cite{Duguet:2025hwi},
\vspace*{-0.2cm}
\begin{align}\label{eq:2body}
\lr{\varepsilon_n^2}=\frac{1}{2A\lr{\hat{R}_n}^2}\left[\lr{\hat{R}_{2n}}+(A-1)\lr{\hat{E}_n}\right]\;,
\end{align}
where $\hat{R}_n=r_{\perp}^n$ acts on one nucleon and $\hat{E}_n(\mathbf{r}_1,\mathbf{r}_2)=r_{1\perp}^{n}r_{2\perp}^{n}e^{in(\phi_1-\phi_2)}$ on two, the expectation values being taken in the nuclear ground state. The first term is what remains when the nucleons are sampled independently from the one-body density, corresponding to $a$ in Eq.~\eqref{eq:4}. The second measures the azimuthal correlation between distinct nucleons, and vanishes if the two-body density factorizes. Evaluating it for a rigid rotor gives $b\beta_n^2$~\cite{Jia:2021qyu,Duguet:2025hwi}, so the quadrupole deformation used throughout this work is a parametrization of the two-body dynamical term.

Two consequences follow for the strategy of this paper. First, both terms are built from the same nucleons of the same colliding nuclei, and on general grounds they should therefore share a longitudinal profile, up to a weak Bjorken-$x$ dependence of the partonic content each nucleon presents to the collision~\cite{Singh:2023rkg}. The universality demonstrated in Fig.~\ref{fig:univ} is a direct test of this expectation in a transport model, for the spherical-baseline and deformation-driven components.

Second, what the subtraction of Eq.~\eqref{eq:5} isolates is a difference in the dynamical term, so any pair of nuclei of similar mass with nearly equal one-body densities but different nucleon correlations would suffice, whether from static deformation, from shape fluctuations, from clustering, or from correlations with no collective interpretation at all. The comparison used here, $^{238}$U versus $^{197}$Au, is the case in which the difference is largest, but it is one member of a wider family. 

\subsection{How the decorrelation map controls $r_2(\eta)$ and bias factor $B_2$}\label{app:I}
The bias factor of Eq.~\eqref{eq:bias} is controlled by the decorrelation, but it is not proportional to the decorrelation strength. To see this, consider the exponential form commonly used to describe decorrelation data~\cite{CMS:2015xmx,ATLAS:2017rij,ATLAS:2020sgl,Nie:2020trj},
\begin{align}\label{eq:c1}
R_2(\eta_1,\eta_2) = \exp\left(-F_2 |\eta_1-\eta_2|\right)\approx 1-F_2 |\eta_1-\eta_2|\;,
\end{align}
For this form the losses of correlation across successive rapidity intervals simply add up, so that $R_2(\etaref,-\etaref)=R_2(\eta,\etaref)\,R_2(\eta,-\etaref)$ whenever the particle of interest lies between the two references, $|\eta|<\etaref$. The two exponents entering the numerator of Eq.~\eqref{eq:bias} then sum to the exponent entering the denominator, and $B_2\equiv1$. The slope $F_2$ may be arbitrarily large and the three-subevent method still returns the true flow. 

\begin{figure}[t]
\includegraphics[width=0.9\linewidth]{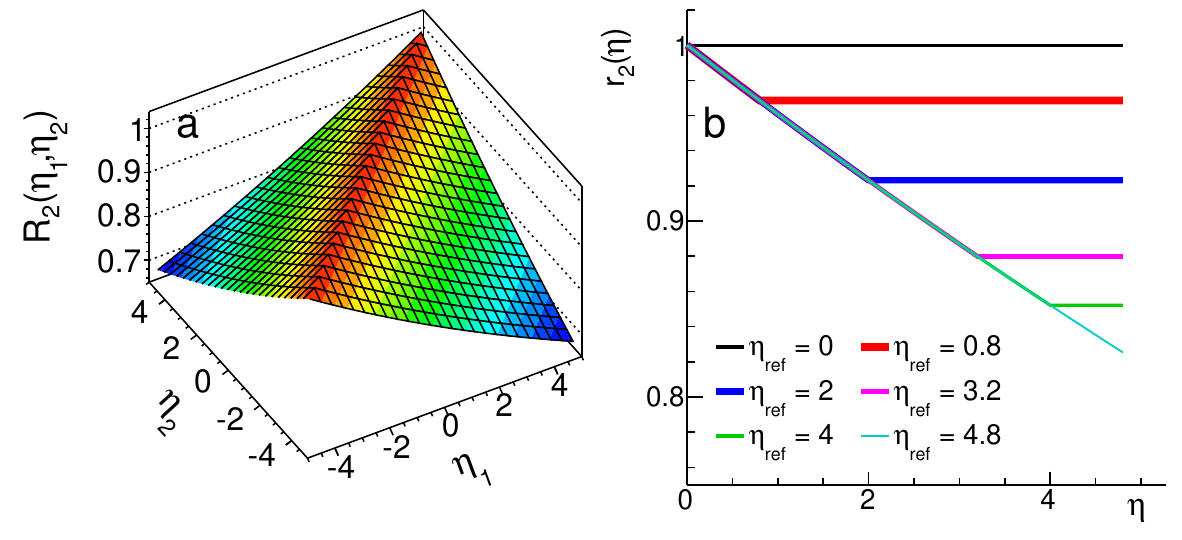}
\caption{\label{fig:bias1b} \textbf{The additive case, based on Eq.~\eqref{eq:c1}.} (\textbf{a}) The map $R_2(\eta_1,\eta_2)$ and (\textbf{b}) the projections $r_2(\eta)$ for six $\etaref$ values, both for the typical slope $F_2=0.02$. The projections follow one common line while $\eta<\etaref$ and saturate beyond it, as Eq.~\eqref{eq:c2} requires. For this model the three-subevent estimator of Eq.~\eqref{eq:bias} gives $B_2=1$. Panel (\textbf{b}) is drawn for $\etaref>0$. Since $r_2$ is unchanged under $\etaref\rightarrow-\etaref$, the panel can be compared directly with Fig.~\ref{fig:proj}b.}
\vspace*{0.25cm}
\includegraphics[width=0.9\linewidth]{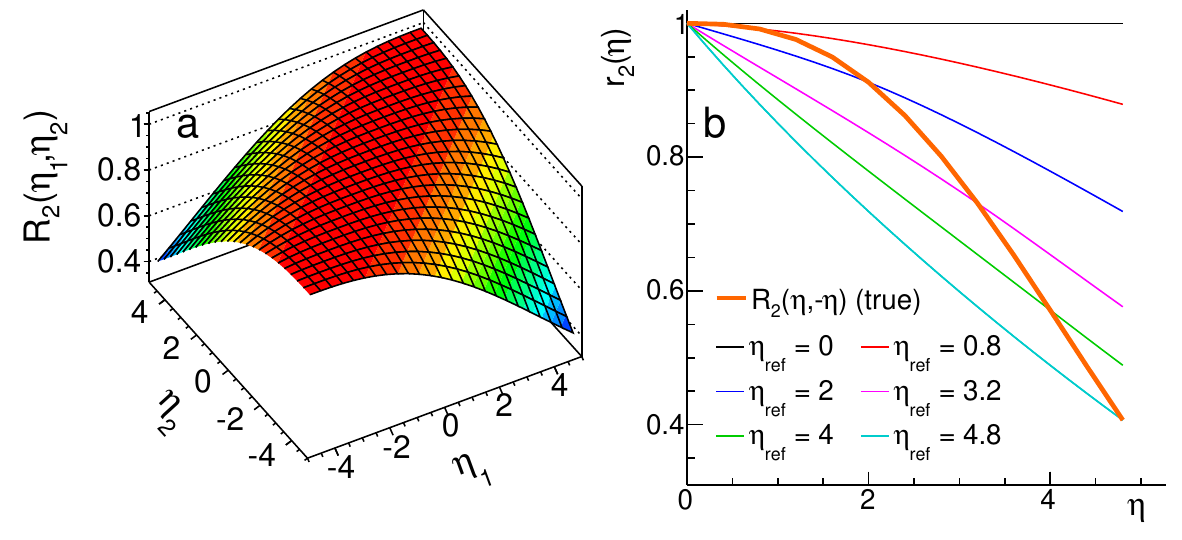}
\caption{\label{fig:bias1c} \textbf{The non-additive case, based on Eq.~\eqref{eq:c4} with $p=2.6$ and $L=10$.} (\textbf{a}) The map $R_2(\eta_1,\eta_2)$ and (\textbf{b}) the projections $r_2(\eta)$ for six $\etaref$ values (thin lines), all of which differ from the true decorrelation $R_2(\eta,-\eta)$ (thick line), meeting it only at $\eta=\etaref$. The transport model projections of Fig.~\ref{fig:proj}b behave in the same way.}
\end{figure}

Figure~\ref{fig:bias1b}a shows this map for $F_2=0.02$, the typical value in central Au+Au collisions~\cite{Nie:2020trj}. A ridge of unit height runs along the diagonal and falls away from it with a fixed slope. The cancellation is exact only when the particle of interest lies between the two references. Outside that range the two intervals in the numerator of Eq.~\eqref{eq:bias} overlap instead of tiling the gap between the references, so together they lose more correlation than the denominator. The bias factor then drops below unity slowly as $B_2=e^{-F_2(|\eta|-\etaref)}$. For $\etaref=4$ it is $0.99$ at $\eta=4.5$ and $0.98$ at $\eta=5$.

The $r_2(\eta)$ of Eq.~\eqref{eq:4b}, evaluated in the quadrant $\eta>0$ and $\etaref>0$, takes the form
\begin{align}\label{eq:c2}
r_2(\eta)=\frac{R_2(-\eta,\etaref)}{R_2(\eta,\etaref)}=e^{-2F_2\min(\eta,\etaref)}\;.
\end{align}
Equation~\eqref{eq:4b} itself puts the reference at negative rapidity, which reverses the sign of $\etaref$ but leaves $r_2$ unchanged. The behaviour of $r_2$ splits at $\eta=\etaref$. Measurements choose $\etaref\gg\eta$ to suppress nonflow, and $r_2$ there falls as $e^{-2F_2\eta}\approx1-2F_2\eta$, the familiar linear form whose slope measures the strength of the decorrelation. However, when $\eta>\etaref$, $r_2$ saturates at the constant $e^{-2F_2\etaref}$. Both behaviours are visible in Fig.~\ref{fig:bias1b}b. The independence of $\etaref$ below the reference is often taken as evidence of genuine decorrelation, and any explicit $\etaref$ dependence is blamed on nonflow contamination.

The same cancellation holds for the wider family
\begin{align}\label{eq:c3}
R_2(\eta_1,\eta_2)=\exp\left[-|G(\eta_1)-G(\eta_2)|\right]\;,
\end{align}
with $G$ any monotone function of rapidity. Here too $B_2\equiv1$ for $|\eta|<\etaref$, because a monotone $G$ is only a change of variable, one that maps this family onto the exponential form by rescaling the $\eta$ axis of Fig.~\ref{fig:bias1b}. The rate at which correlation is lost may therefore vary arbitrarily with rapidity, yet the bias still vanishes.

A two-parameter family that breaks the additive rule is 
\begin{align}\label{eq:c4}
R_2=\exp[-(|\Deta|/L)^{p}]\;,
\end{align}
where $L$ sets a rapidity range and $p$ is an exponent. Inserting this form into Eq.~\eqref{eq:bias} gives $B_2(0,\etaref)=\exp[(\etaref/L)^{p}(2^{\,p-1}-1)]$. The sign of the bias is therefore fixed by the factor $2^{\,p-1}-1$ alone. For $p<1$ that factor is negative and the bias factor falls below unity. It vanishes at $p=1$, where $B_2=1$. For $p>1$ it is positive and the bias factor rises above unity. At $p=1$ the form reduces to Eq.~\eqref{eq:c1}, and Fig.~\ref{fig:bias1b} corresponds to $p=1$ and $L=50$.

Both parameters have a direct physical meaning. $L$ is the largest rapidity separation that a single source can still correlate, for example the length of a string, and $p$ says how abruptly the correlation is lost once that separation is exceeded. Strings do not all have the same length, however, so what the $R_2$ map sees is a distribution of lengths. When the lengths cluster around a common value, the correlation survives out to that value and then falls away quickly. This is the superadditive case, with $p>1$ and $B_2>1$. A wide spread of lengths gives the subadditive case, with $p<1$ and $B_2<1$, because the shortest strings drop out first and the correlation then falls quickly at small separations and only slowly afterwards. A superposition of exponentials is an example of this. 

Figure~\ref{fig:bias1c} shows the same family for $p=2.6$ and $L=10$. This choice approximately reproduces the transport model map of Fig.~\ref{fig:v2d}d and the projections of Fig.~\ref{fig:proj}b. The map in Fig.~\ref{fig:bias1c}a has a flat plateau along $\eta_1=\eta_2$ and then falls rapidly. Such a shape is expected when the sources of correlation have a finite longitudinal extent, for example strings that span about two units in $\eta$ and into which energy is deposited~\cite{Jia:2024xvl}. The projections in Fig.~\ref{fig:bias1c}b behave quite differently from those of Fig.~\ref{fig:bias1b}. They no longer collapse onto a common line for $|\eta|<|\etaref|$, and they no longer flatten beyond it, so both signatures of the additive case are lost. Since $r_{2,\beta}$ is built from $V_{2\Delta,\beta}$, from which Eq.~\eqref{eq:5} has already removed the nonflow, this $\etaref$ dependence cannot come from residual short-range correlations. It is nevertheless the kind of dependence that nonflow contamination produces, so without the subtraction the two would be hard to tell apart in data.

\subsection{What the decorrelation map determines about the initial geometry}\label{app:J}
The map $R_2(\eta_1,\eta_2)$ is a second moment of the event-by-event distribution of $V_2(\eta)$, and it is worth saying how much of that distribution it fixes.

In each event $V_2(\eta)$ is a complex function of rapidity with an amplitude and an orientation, and the ensemble is the distribution of those functions. The correlator $V_{2\Delta}(\eta_1,\eta_2)$ is the covariance of that ensemble, so it can be written as $V_{2\Delta}(\eta_1,\eta_2)=\sum_{\alpha}\lambda_{\alpha}\psi_{\alpha}(\eta_1)\psi_{\alpha}^{*}(\eta_2)$ with orthogonal modes $\psi_{\alpha}(\eta)$ and positive weights $\lambda_{\alpha}$~\cite{Bhalerao:2014PCA}. A single mode gives $R_2\equiv1$ and no decorrelation at all, because the rapidity dependence then cancels between the numerator and the denominator of Eq.~\eqref{eq:1}. Decorrelation is the statement that more than one mode is present, and the shape of the map is sensitive to how many there are and what rapidity profiles they have. The same decomposition has been applied to the transverse momentum dependence of $V_n$~\cite{CMS:2017PCA,Mazeliauskas:2015vea,Mazeliauskas:2015efa}, where the subleading mode carries a node.

The nonflow removal introduced in this work is critical for extracting the correct geometry-driven decorrelations. In the transverse momentum case a mixture of nonflow with the leading flow mode can generate a non-collective subleading mode whose magnitude may exceed the genuine one~\cite{Liu:2020ely}. The same mechanism can operate along rapidity, so a mode decomposition is only meaningful on a map from which the nonflow has been removed. That is what the system comparison of Eq.~\eqref{eq:5} provides.

Once the map fixes the statistics of $V_2(\eta)$, the linear response $V_2(\eta)\propto\mathcal{E}_2(\eta)$ carries that information to the initial geometry. The normalized map built from $V_2(\eta)$ would then give the same information as the one built from $\mathcal{E}_2(\eta)$, because a response factor common to all events cancels in the ratio. This last step completes the imaging of the geometry of the fireball.

How realistic that image is depends on the nature of the fluctuations. If the ensemble of $V_2(\eta)$ is Gaussian, the covariance determines the whole distribution, and the $R_2(\eta_1,\eta_2)$ map then fixes the rapidity correlation structure of $\mathcal{E}_2(\eta)$ completely. Departures from Gaussianity would have to be reached through multiparticle decorrelations.

\subsection{Decorrelation bias in four-subevent methods}\label{app:K}
In practice, the three-subevent method of Eq.~\eqref{eq:bias} may not be feasible or optimal for three reasons. First, the detector may lack good acceptance at both forward and backward rapidities. Second, the available acceptance may not provide a sufficient rapidity gap to suppress nonflow. Third, in asymmetric systems such as $p/d/^3$He+Au collisions, the flow signal is small and the nonflow is large in the light-ion-going direction. In such cases, a four-subevent method can be used to alleviate the nonflow issue, with the flow signal measured as
\begin{align}\label{eq:bias1}
v_2(\pT)_{\mathrm{est.}} &=\frac{\lr{V_2(\pT,\eta_1) V_2^*(\eta_2)}}{\sqrt{\frac{\lr{V_2(\eta_3) V_2^*(\eta_2)}  \lr{V_2(\eta_4) V_2^*(\eta_2)}}{\lr{V_2(\eta_3) V_2^*(\eta_4)} }}}\\\nonumber
&=v_2(\pT,\eta_1)\, B_{2,\mathrm{4sub}}\\\label{eq:bias2}
B_{2,\mathrm{4sub}} &= R_2(\eta_1,\eta_2)\sqrt{\frac{R_2(\eta_3,\eta_4)}{R_2(\eta_3,\eta_2)\,R_2(\eta_4,\eta_2)}}\;.
\end{align}
\begin{figure}[t]
\includegraphics[width=0.95\linewidth]{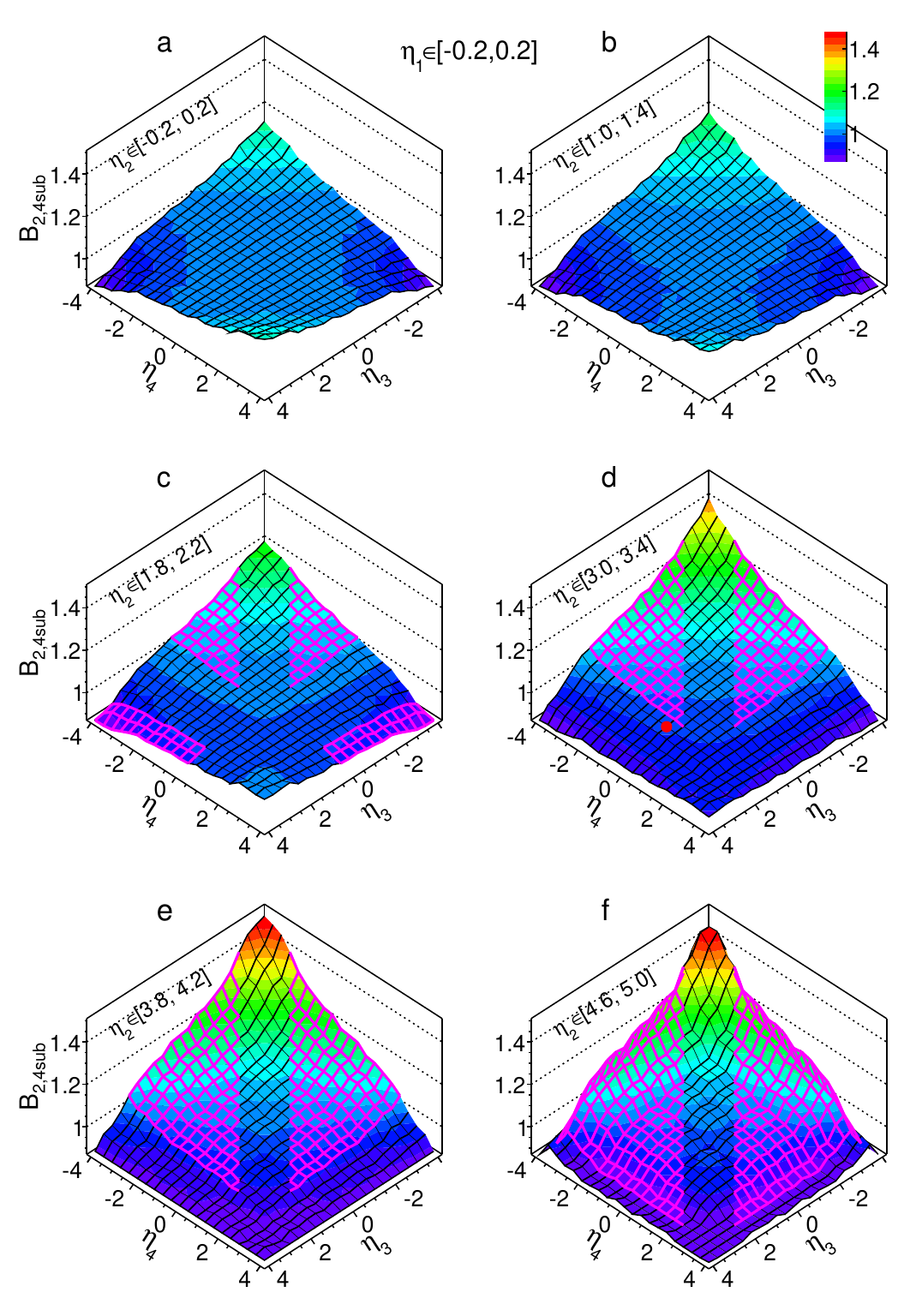}
\caption{\label{fig:bias2} \textbf{Decorrelation bias in the four-subevent method.} The bias factor $B_{2,\mathrm{4sub}}$ of Eq.~\eqref{eq:bias2} for $v_2(\pT)$ measured at mid-rapidity ($|\eta_1|<0.2$), shown as a function of the auxiliary rapidities $\eta_3$ and $\eta_4$ for six choices of the reference rapidity $\eta_2$ (panels \textbf{a}--\textbf{f}), calculated from $R_{2,\beta}(\eta_1,\eta_2)$ in Fig.~\ref{fig:v2d}d. The magenta mesh marks the region satisfying a minimum $\eta$ gap of one unit for all two-particle pairs in Eq.~\eqref{eq:bias1}. The red circle in panel \textbf{d} marks the centroid of the region used in the previous PHENIX analysis~\cite{PHENIX:2018lia}.}
\end{figure}
This method, also referred to as the $3\times$2PC method, is often used in small-system analyses~\cite{PHENIX:2017nae,PHENIX:2018lia,PHENIX:2021ubk,STAR:2023wmd}. For example, one may choose $\eta_1=\eta_4\sim0$, $\eta_2\sim2$, and $\eta_3\sim4$. This ensures at least a two-unit $\Deta$ gap for every two-particle correlator in Eq.~\eqref{eq:bias1}. The POI rapidity is used for both $\eta_1$ and $\eta_4$. This is permissible because the two never appear in the same pair. A concrete example is the PHENIX analysis of $p/d/^3$He+Au collisions, which used $|\eta_1|,|\eta_4|<0.35$, $\eta_2\in[-4,-3]$, and $\eta_3\in[-3,-1]$~\cite{PHENIX:2018lia}. This choice may retain some nonflow in $\lr{V_2(\eta_3)V_2^*(\eta_2)}$, but the contamination is more manageable and, unlike in the three-subevent method, does not enter in a $\pT$-dependent way.

Because the decorrelation depends on the rapidities separately and is strongly system dependent, each choice of $(\eta_1,\eta_2,\eta_3,\eta_4)$ carries its own bias factor $B_{2,\mathrm{4sub}}$. However, once the full $R_2(\eta_1,\eta_2)$ map is known, $B_{2,\mathrm{4sub}}$ can be predicted over the entire phase space, and the combinations that minimize the bias can be identified. Figure~\ref{fig:bias2} shows such maps, calculated from $R_{2,\beta}(\eta_1,\eta_2)$ of Fig.~\ref{fig:v2d}d. We fix the POI at mid-rapidity, $|\eta_1|<0.2$, and each of the six panels takes one value of the reference rapidity $\eta_2$. The two axes within a panel are the auxiliary rapidities $\eta_3$ and $\eta_4$, so every point of a panel is one possible four-subevent configuration, and the value plotted there is its bias factor.

The bias is small when $\eta_2$ lies close to the POI, but that choice is not usable, because the correlator $\lr{V_2(\pT,\eta_1)V_2^*(\eta_2)}$ picks up a large and $\pT$-dependent nonflow contamination. A better choice is $|\eta_2|\gtrsim2$, which the four panels of Fig.~\ref{fig:bias2}c--f satisfy. The two members of every pair in Eq.~\eqref{eq:bias1} should also be separated by at least one unit of rapidity to suppress nonflow. These requirements leave only the regions marked by the magenta mesh. Even inside those regions the bias factor $B_{2,\mathrm{4sub}}$ may deviate significantly from unity, ranging from about 0.8 to 1.4. The centroid of the region used in the PHENIX analysis~\cite{PHENIX:2018lia}, marked by the red circle, lies slightly outside this nonflow-safe region. The corresponding bias factor is $B_{2,\mathrm{4sub}}=0.96$, i.e., a 4\% bias. The default PHENIX combination therefore happens to carry only a small decorrelation bias. Decorrelations are expected to be considerably stronger in the small systems themselves, and the real bias there can only come from their own $R_2$ maps. Those maps can now be measured with the comparison method introduced in this work.

The small-system puzzle at RHIC shows what is at stake. Across $p$+Au, $d$+Au and $^3$He+Au collisions the two experiments obtain conflicting $v_3$ orderings~\cite{PHENIX:2018lia,STAR:2023wmd}. PHENIX correlates mid-rapidity particles with backward, Au-going references, whereas STAR correlates particles closer to mid-rapidity. Model studies attribute about half of the difference to longitudinal decorrelations sampled differently by the two acceptances~\cite{Zhao:2022ugy}. In the present language the two acceptances are simply two points on a bias map such as Fig.~\ref{fig:bias2}. Once $R_2(\eta_1,\eta_2)$, and its triangular counterpart $R_3(\eta_1,\eta_2)$, are measured free of nonflow, the bias factor of each detector combination can be calculated, and $v_n$ from different acceptances can then be compared on a common footing. 

The required maps $R_2$ and $R_3$ can be extracted with the nonflow subtraction introduced in this work, using pairs of systems with similar multiplicity but different geometry. Suitable pairs are $^{16}$O+$^{16}$O versus $^{20}$Ne+$^{20}$Ne at the LHC~\cite{Giacalone:2024luz}, and $d$+Au versus $^{16}$O+$^{16}$O at RHIC. The resulting bias maps, analogous to Fig.~\ref{fig:bias2}, will show which rapidity combinations minimize both the nonflow and the decorrelation bias. 

\bibliography{ref}{}
\bibliographystyle{apsrev4-1}
\end{document}